\documentclass[journal]{IEEEtran}
\usepackage{amsmath,amsfonts, amsfonts}
\usepackage{algorithmic}
\usepackage{algorithm}
\usepackage{array}
\usepackage[caption=false,font=normalsize,labelfont=sf,textfont=sf]{subfig}
\usepackage{textcomp}
\usepackage{stfloats}
\usepackage{url}
\usepackage{verbatim}
\usepackage{graphicx}
\usepackage{cite}
\usepackage{xcolor}
\newtheorem{definition}{Definition}

\begin{document}

\title{Seizure detection from Electroencephalogram \\signals via Wavelets and Graph Theory metrics}

\author{Paul Grant, 
 Md Zahidul Islam

\thanks{P Grant and Z. Islam are with the School of Computing, Mathematics and Engineering, Faculty of Business, Justice \& Behavioural Sciences at Charles Sturt University, Bathurst, NSW Australia. Email:(pgrant, zislam)@csu.edu.au }
\thanks{Corresponding Author: Paul Grant. }
}

\markboth{
}{Shell \MakeLowercase{\textit{et al.}}: Seizure detection from Electroencephalogram signals}


\maketitle

\begin{abstract}
Epilepsy is one of the most prevalent neurological conditions, where an epileptic seizure is a transient occurrence  due to abnormal, excessive and synchronous activity in the brain. Electroencephalogram signals emanating from the brain may be captured, analysed  and then play a  significant role in detection and prediction of epileptic seizures. 
Here we apply 
 the Maximum Overlap Discrete Wavelet Transform to both reduce signal \textit{noise} and use signal variance exhibited at differing inherent frequency levels to develop various metrics of connection between the electrodes placed upon the scalp. 
 Using short duration epochs, to approximate close to real time monitoring, together with simple statistical parameters derived from the reconstructed noise reduced signals we initiate seizure detection. To further improve performance we utilise  graph theoretic
 indicators from derived electrode connectivity. From there we  build the attribute space.  We utilise  open-source software and publicly available data to highlight the superior Recall/Sensitivity performance 
 of our approach, when compared to existing published methods.
\end{abstract}

\begin{IEEEkeywords}
Graph Theory Signal analysis, Wavelet transforms.
\end{IEEEkeywords}

\section{Introduction}
\label{sec:Intro}
The role of electroencephalogram (EEG) in the diagnosis and classification of seizure
types and epilepsy syndromes is well-established.
Information regarding brain activity is extensively collated via EEG signals where differing brain activities  may be studied. EEG signals provide  information on nerve cell activity in the brain.. Such signals may be used in classification and characterisation of seizures
and seizure syndromes and to support the clinical diagnosis of a
seizure, epilepsy, or epilepsy syndrome \cite{P1Tatum}.  When neuronal activity is sufficiently disturbed, seizures may result. A common problem is classification of such signals into their respective stages, i.e. pre-seizure, 
 seizure and post-seizure. 
A seizure is a transient irregularity in the brain's electrical processes that may also produce disruptive physical symptoms \cite{P1Zandi}.
Such signals may be  collected from the surface of the scalp, via electrodes which may be  situated in defined spatial arrangements. One such arrangement is the ``10-20'' system of electrode placement.

The EEG frequency spectrum contains  five medically established EEG rhythms, i.e. delta
(0–4 Hz), theta (4–8 Hz), alpha (8–13 Hz), beta (13–30
Hz) and gamma (30–60 Hz or 30–100 Hz) \cite{P3Tsip:2019 ,P4aKumar:2012}. 
\noindent The EEG measures both the frequency and amplitude of electrical activity generated from the brain. 
 EEG is complex nonlinear, non-periodic, and non-stationary data containing a large amount of information \cite{Xin:2022}. It has been noted that highly non-stationary signals in biomedicine may benefit from Time frequency tools, i.e. EEG analysis in epilepsy patients \cite{Stank:2013}. Visual inspection of the EEG signals is time-consuming and not a trivial task to undertake and monitor in real time, requiring concentration as well as observation across many outputs. Inpatient  EEG telemetry is expensive, labour-intensive to collate and a limited resource.
Even then long term EEG monitoring is unlikely to be
productive if the patient’s events occur less than once per
week \cite{P5Smith:2005}. 
Research on EEG analysis algorithms and automatic
detection for epilepsy began in the 1970s \cite{Xin:2022}.
Many automated EEG signal classification and seizure
detection systems, using different approaches, have
emerged in recent years \cite{P7Gajic:2014}.

Feature extraction from the raw EEG data is generally categorised as either time-domain or frequency-domain. Time-domain features are easily computed, e.g. sum, average,  standard deviation and energy, all derived from the signal's values over defined time periods. Frequency-domain features are most often obtained by transposing  the EEG signals into specific frequency bandwidths related to the problem at hand.
   Many such methods extracting features from the time and/or frequency domains are also mentioned \cite{P8Lee:2014}. 
   
   For EEG classification, ensemble methods have attracted attention \cite{P9Abua:2015}.  It has been noted that the capability of these ensemble methods is subject to not only the type of base classifier but also the settings and parameters  used for each individual classifier. Generally speaking, ensemble learning for EEG signal classification is effective, however different ensemble methods may return varying levels of performance and may even deteriorate if using unsuitable parameters
\cite{P10Sun:2007}.

\subsection{Related work}
Time and frequency domains are commonly used in feature extraction methods for EEG signals. To analyse the frequency domain features, the Discrete Fourier Transform (DFT) has been applied \cite{P8Lee:2014}. Decomposing a signal in terms of its frequency content, using the DFT which relies on  sinusoids, results in fine resolution within the frequency domain. For Fourier analysis the idea is, given a time series $X_t$ see Definition \ref{Def:series1}, one obtains \begin{equation}
X_t =  \sum_k \{A_k sin(2\pi f_kt) + B_k cos(2\pi f_kt)
\end{equation}
where  $f_k$ represents a collection of different frequencies. 

\begin{definition}\label{Def:series1}

Let a sequence $ X_1, X_2, \ldots , X_n $ represent  a time series of $n$ elements, denoted as $\{X_t : t = 1, \ldots, n\},\; X_t  \in \mathbb{R}$.
\end{definition}

However the Fourier series representation is not very effective at  time resolution \cite{P11Vika:2015}, all the transform provides is which frequency components are present in the signal.  When the time localisation of the frequency components are required, a transform providing a time frequency representation of the signal is needed.  One such method of  decomposing signals is the  Wavelet Transform (WT), where a  wavelet  is a function such that $\psi = L^2(R)$  with a zero average,  $\int_{-\infty}^{+\infty} \psi(t)dt = 0$.   We may define the Continuous Wavelet Transform (CWT) of a signal or a time series, $X_t$  as 
\begin{equation}\label{CWT}
CWT_\psi X(a,b) = \frac{1}{\sqrt{|a|}}\int_{-\infty}^{+\infty} X_t \psi^* \left(\frac{t-b}{a}\right)dt 
\end{equation} where $\psi(t)$ is called the mother wavelet, with the asterisk denoting the complex conjugate. 
The parameter $a$ determines the frequency of oscillation and length of the wavelet while parameter $b$ is the shift of the wavelet in time \cite{P12Omer:2012}.

Wavelets represent the signal in terms of functions that are localised both in time and frequency, a time-frequency approach.  As we may not know which frequency component exists at a given time instant, all we may do is consider which components exist  at a specified time interval.  Wavelets provide a variable resolution, higher frequencies are better resolved in time, and lower frequencies are better resolved in frequency. This means that, a certain high frequency component can be located better in time (with less relative error) than a low frequency component. On the contrary, a low frequency component can be located better in frequency which results in  good time resolution at high frequencies, and good frequency resolution at low frequencies \cite{P13Poli:1994}.
Time-domain features have been used to extract information from EEG waves \cite{P14Sidd:2018}, in that work data from 76 electrodes was used, altering the epoch duration and deriving nine statistical parameters for each signal per epoch. The aim was to discover if accuracy may be retained as the duration of the epoch was reduced as well as using a decision forest classifier in an attempt to discover useful knowledge discovery as opposed to a black box approach.  
No consideration of the specific relevant frequency bandwidths was considered. Therefore as all inherent frequencies in the original signal were included in the  analysis, resulting in  the possibility of noise\footnote{Noise here refers to  information (\textit{energy}) within frequency bandwidths not related to the event under consideration.} being included in such analysis.

In an attempt to consider the established frequencies considered as EEG rhythms, 
EEG spectral analysis has frequently been undertaken  using wavelet transforms \cite{Zach:2013}. A standard approach here is to decompose the signal into frequency sub-bands and then derive parameters from the resulting wavelet coefficient values \cite{P15Jaco:2018}, \cite{ P16Chen:2017}, \cite{ P17Suba:2007}. The most commonly used form of the wavelet transform is the Discrete Wavelet transform (\textbf{DWT}), see Section \ref{subs:dwt}. The DWT is defined by using discrete values of the parameters $a\, \&\, b$ from (\ref{CWT}). \par
      Shrestha, Dahi Shrestha and Thapa  \cite{P18Shre:2019}  using the DWT, determined which frequency bands provided an optimal classification outcome from only one individual. Then used those bandwidths for feature extraction across other individuals. Only data from eight electrodes were used, as opposed to 23 available to them. They overlapped and windowed the data to effectively double the amount of data available. This may hinder real time monitoring. A black box approach to classification was undertaken using an Artificial Neural Network (ANN).
     While the DWT returns the same number of coefficients as in the original time series, the DWT provides a reducing number of coefficients within each decomposition level as the level increases, which may hinder development of derived indicators at the higher decomposition levels.
    Another  variant  of the wavelet transform is  Maximum Overlap Discrete Wavelet Transform (\textbf{MODWT}). The MODWT returns a number of coefficients which are considerably more numerous than the number of data points in the original signal. \par 
    The application of graph theoretic principles to EEG analysis may be noted \cite{P19Cao:2013, P20Subr:2013}.  What is not common is consideration of the topology of EEG signal channels, so as to not overlook important spatial information \cite{P20Li:2021}. Also the determination method of functional connectivity (FC) between the various electrodes or node points, thats allows one to determine simulated links and hence derive graph theoretic indicators. Several FC  methods are possible, the choice of which is not trivial \cite{P21Fall:2014}. 
   The  variance of MODWT wavelet coefficients has been previously used  to  construct such a FC \cite{P22Acha:2007}. There the motivation was the adoption of the  relatively novel \textit{metrics of network efficiency}.  However that study was concerned with using magnetic resonance imaging, which included approx 90 node points,  to detect the affect of an induced drug on brain behaviour.
    \subsection{Main contributions of paper}
   Using the DWT regarding EEG analysis has become quite common as it has some advantages in analysis of time series i.e. Time and frequency information, data segmented into various components and hence easier to eliminate unwanted portions and flexible choice of DWT Bases.
   Wavelet-based estimators of the
correlation between two processes have been
shown to have desirable statistical properties \cite{P24Gutt:2000}. Here
    we utilise the MODWT to decompose a   signal, reconstruct a noise reduced signal as well as determine functional connectivity between specific  areas and hence develop graph (\textit{"small world"} \cite{P29Lato: 2001}) theoretic indicators  at different wavelet scales or defined frequency sub-intervals.  This permits monitoring of several areas/lobes of the brain  and construction of a topological representation of interconnectivity, across a network derived from a small set of EEG electrodes/nodes  during 1 second epochs, which would permit close to real time monitoring. 
We utilise readily  derived statistical parameters from a signal restricted to the relevant EEG rhythms and implement graph theoretic metrics, some of which are distance weighted.  These parameters have shown to differ significantly during the seizure or non-seizure states \cite{23Gran:2021}.
\section{Preliminaries}
We introduce an overview of the main transform method and the graph theoretic metrics that we implement when applying our method.
\subsection{Wavelets}
Wavelets are functions that segment the  data into different frequency components,
where it is possible to then study such components at the resolution level that are matched to its scale \cite{P25Grant:2019}.
 Wavelets have been applied in many areas
such as signal processing, image denoising, data compression, speech recognition and quantum mechanics.
If we represent a function by wavelets, such a representation is called a wavelet transform. The wavelet transform has good local
time-frequency characteristics and can describe the signal
features well in both time and frequency domains \cite{Xin:2022}. If we use a wavelet that is discretely sampled, resulting in a wavelet filter which we apply to the initial data, then we have DWT. The output resulting from the DWT is a data vector being the  same length as that of the initial input.
\subsubsection{DWT} \label{subs:dwt}
Given a  sequence $X_t$ as per Definition \ref{Def:series1} where $n = 2^J\; :\;J \in Z^+$, the DWT is a linear transform of $X$ producing $n$ DWT coefficients. In vector notation with 
the DWT gives $\textbf{W} = \mathcal{W}X$ where
\begin{itemize}
\item $\textbf{W}$ is a vector of $J + 1$  levels and $n$ DWT coefficients
\item $\mathcal{W}$ is a \textit{$n \times n$ orthonormal} transform matrix.

\end{itemize}

\begin{equation}    \textbf{W} = 
     \begin{bmatrix}
         W_1\\
         W_2\\
         W_3\\
         V_3
     \end{bmatrix} 
      =
     \begin{bmatrix}
     \mathcal{W}\\
     \end{bmatrix}
     X
     \label{eqn:eq1}
\end{equation}

DWT transforms $X$ into 2 different types of coefficients; 
 $W_i$ being the detail wavelet coefficients with $i:1, \ldots, J$. and  $V_J$ the smooth coefficients.  
 

Equation \ref{eqn:eq1} permits  $X$ to be expressed as the sum of $J+1$ vectors; 

\begin{equation} \label{DWTadd}
 X_t= \sum^J_{j=1} \mathcal{W}^T_j\textbf{W}_j + \mathcal{V}^T_J\textbf{V}_J  \equiv \sum^J_{j=1} \mathcal{D}_j + \mathcal{S}_J  
\end{equation}
Equation \ref{DWTadd}  represents a Multiresolution analysis (\textbf{MRA}) with $\mathcal{D}_j$ being the detail coefficients at scale $j$ and $\mathcal{S}_J$ being the smooth coefficients at scale $J$.
The construction of the DWT has been well-defined \cite[Ch. 4]{P26Perc:2000}.


\subsubsection{MODWT}
The MODWT is constructed in an entirely similar manner to the DWT  and offering similar properties. It is  sometimes known as the undecimated DWT or stationary DWT. The important
differences of MODWT from DWT are that it is a highly
redundant and non-orthogonal transformation. This transform provides the same number of wavelet coefficients at each decomposition level. The additional number of wavelet coefficients increases the degree of freedom and reduces the variance of the estimates \cite{P26aCand:2020}.\par  MODWT is achieved by using a different transform matrix $\mathcal{\widetilde{W}}$ to that in  (\ref{eqn:eq1}), therefore we alter the transform and not the signal. The MODWT coefficients may be seen as the differences of weighted averages from the original observations. The MODWT is defined for all sample sizes $n$ and when moving to  each resolution scale, 
we do not downsample as in the DWT.  The size of each of  the vectors resulting from the MODWT, $\, \widetilde{W_1}, \widetilde{W_2}, \ldots , \widetilde{W_{J_0}}, \widetilde{V_{J_0}}$ is $n$. Therefore a MODWT to level $J_0$ returns $(J_0 + 1)n$ values.

For $J_0$ level MODWT based MRA, see  (\ref{eqn:MODWTaddi}) which is analogous to the DWT based MRA. 
\begin{equation} \label{eqn:MODWTaddi}
 X_t = \sum^{J_0}_{j=1} \mathcal{\widetilde{W}}^T_j\widetilde{\textbf{W}}_j + \mathcal{\widetilde{V}}^T_{J_0}\widetilde{\textbf{V}}_{J_0} \equiv  \sum^{J_0}_{j=1} \mathcal{\widetilde{D}}_j + \mathcal{\widetilde{S}}_{J_0}  
\end{equation}

For a complete overview and construction of the MODWT as well as advantages over the DWT see \cite[Ch. 5]{P26Perc:2000}.

\subsubsection{Wavelet variance}\label{subs:wavar}
While the application of Wavelets within EEG analysis is often encountered, the use of the variance within the wavelet coefficients resulting from the transform, mainly consider this wavelet variance as a selected feature only. Usually this is  only the standard deviation of the detail wavelet coefficients at  each decomposition scale or a linear combination of the variance  across the different wavelet decomposition scales \cite{P27Asho:2017, P28Janj:2010}. Little if any consideration is given  to the covariance between different signals/nodes, which may be calculated similarly as the wavelet variance, this is called wavelet covariance. This wavelet covariance and correlation can be used to derive electrode/node connectivity during an epoch and hence construct a graph or network, from which to provide additional features for classification, especially if the connectivity between electrodes/nodes alters significantly between the non-seizure and seizure states.

From a sequence $X_t$, see Definition \ref{Def:series1}, we are able to calculate the usual parameters: \newline
sample mean : $\bar{X} = (1/n) \sum _{t=1}^{n}X_t $ and \newline sample variance :  $\hat{\sigma}^2_X \equiv (1/n) \sum _{t=1}^{n} (X_t - \bar{X})^2$.

\begin{definition} \label{sam:covar}
let  cov\{$X_t\, , Y_t$\} denote the  covariance of two random sequences $X_t\; \text{and}\;  Y_t$
\begin{equation}
cov\{X_t , Y_t\} = \sum_{t=1}^n \frac{(X_t -\bar{X})(Y_t - \bar{Y}) }{n}
\end{equation}
\end{definition}
Similarly, we may derive the same parameters  for the  coefficients resulting from a wavelet transform. From the MODWT based MRA, see (\ref{eqn:MODWTaddi}), we may derive a scale based energy decomposition.
\begin{equation}\label{MODWT:Anova}
 \parallel X_t \parallel^2 = \sum^J_{j=1} \parallel\widetilde{\textbf{W}}_j\parallel^2 + \parallel\widetilde{\textbf{V}}_J\parallel^2
\end{equation}

This leads to a scale based decomposition of the sample variance:
\begin{equation}
\hat{\sigma}^2_X = \frac{1}{n} \sum_{j=1}^J  \parallel\widetilde{\textbf{W}}_j\parallel^2  + \frac{1}{n}\parallel\widetilde{\textbf{V}}_J\parallel^2 - \bar{X})^2 \equiv \sum_{j=1}^J \nu_X^2(\mathcal{T}_j)
\end{equation}
 Wavelet variance for scale $\mathcal{T}_j$ may be defined as  
\begin{equation}\label{wavVar}
\nu^2_{X}(\mathcal{T}_j) \equiv \text{var}\{\widetilde{W}_j\}
\end{equation}

Given sequences  $X_t \, \& \, Y_t$ as per Definition \ref{Def:series1}, with wavelet coefficients 
$\{\widetilde{W}_{X,j} \}$  and $\{\widetilde{W}_{Y,j} \}$, we have 
wavelet covariance 
\begin{equation} \label{wavCoVar}
\nu_{XY}(\mathcal{T}_j) \equiv cov \{\widetilde{W}_{X,j},\widetilde{W}_{Y,j} \}
\end{equation}
and a decomposition at scale $\mathcal{T}_j$  of the covariance between $\{X_t\}$ and $\{Y_t\}$; 
\begin{equation}\label{covarXY}
\sum_{j=1}^\infty \nu_{XY}(\mathcal{T}_j) =  cov \{X_t\,, Y_t\}
\end{equation} 

From Wavelet covariance we obtain  wavelet correlation \cite[Ch. 8]{P26Perc:2000}.
\begin{equation}\label{wave:corel}
\rho_{XY} \equiv \frac{cov\{\widetilde{W}_{X,j}\,, \widetilde{W}_{Y,j}\}}{(var\{\widetilde{W}_{X,j}\}\, var\{\widetilde{W}_{Y,j}\})^{1/2}} =  \frac{\nu_{XY}(\mathcal{T}_j)}{\nu_X(\mathcal{T}_j)\nu_Y(\mathcal{T}_j)}
\end{equation}
 Using the these equations to derive connectivity across the electrodes, we may use Graph Theoretic methods  to derive additional attributes. 
  \subsection{Graph Theory metrics} \label{subsec:GrpT}
\begin{figure}[!t]
\centering
\includegraphics[width=3.0in]{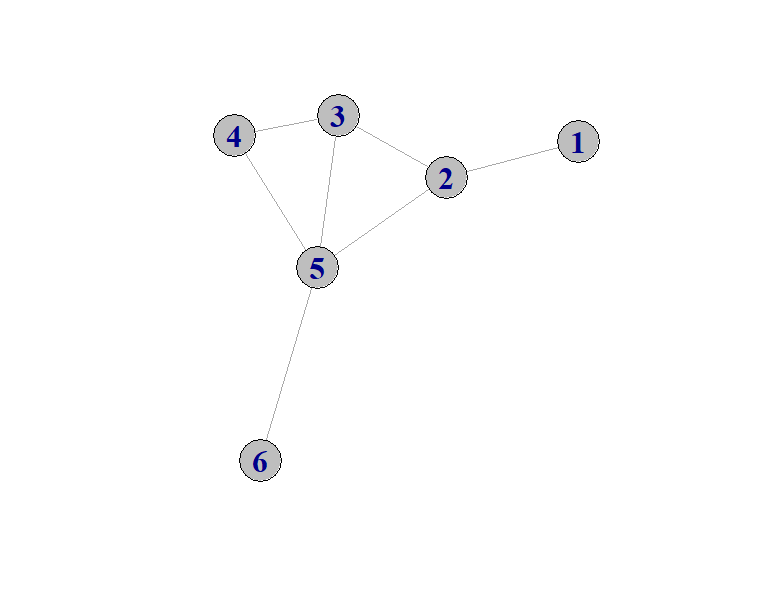}
\caption{Example Graph, a network of connected nodes.\label{fig:examgph}}
\end{figure}

\begin{definition}
A graph is an abstract representation of a set of objects where some pairs of the objects are connected by links.
\end{definition}
 Using Fig. \ref{fig:examgph} as a sample network, which could be interpreted as a representation of connections between the EEG electrodes during an epoch,  it is possible  to develop various parameters which may be used to construct attributes. 

 \subsubsection{Connections}
  From this example network and its inherent connections between the nodes we form an adjacency matrix Table \ref{tab:ExAdl}, where a  connection between two nodes is  represented by a One  and no connection, a Zero.
    The number of connections per node to other nodes may simply be observed as the sum of a row in the adjacency matrix i.e. Node 2 is directly connected to 3 other nodes, therefore degree of connection is equal to 3.

  \subsubsection{Efficiency}
 It is possible to derive how “efficiently”\footnote{Efficiency of $\text{node}_i$  is related  to the sum of the inverses of the minimum distances between $\text{node}_i$ and other $\text{node}_{j \ne i}$ in the network.} the nodes are able to  communicate across their network of connections.  

  \begin{table*}[t!] 
\centering

\caption{Adjacency matrix formed from Fig. \ref{fig:examgph} where ``1'' indicates a connection between the nodes}

\resizebox{0.85\textwidth}{!}{%
 \begin{tabular}{||c c c c c c c || c ||} 
 \hline
  & Node-\textit{1} & Node-\textit{2} & Node-\textit{3} & Node-\textit{4} & Node-\textit{5} & Node-\textit{6} &\textit{Sum}\\ [0.2ex] 
 \hline\hline
 Node-\textit{1} & 0 & 1 & 0 & 0 & 0 & 0  &\textit{1}\\ 
 \hline
 Node-\textit{2} & 1 & 0 & 1 & 0 & 1 & 0  &\textit{3}\\
 \hline
 Node-\textit{3} & 0 & 1 & 0 & 0 & 1 & 1 &\textit{3}\\
 \hline
 Node-\textit{4} & 0 & 0 & 1 & 0 & 1 & 0 &\textit{2}\\
 \hline
 Node-\textit{5} & 0 & 1 & 1 & 1 & 0 & 1 &\textit{4}\\  
 \hline
 Node-\textit{6} & 0 & 0 & 0 & 0 & 1 & 0 &\textit{1}\\  
 \hline
\end{tabular} }
\label{tab:ExAdl}
\end{table*}

 Global efficiency for a node is defined as the inverse of the harmonic mean of the minimum path length between $\text{node}_i$ and all other  $\text{nodes}_{j \ne i}$ in the graph \cite{P29Lato: 2001}.   We may calculate the global efficiency ($Eglob$) for each node, for simplicity we may set the distance from one node to another connected by a single direct link with the length of the link as 1.

 \begin{figure}[!t]
\centering
\includegraphics[width=3.0in]{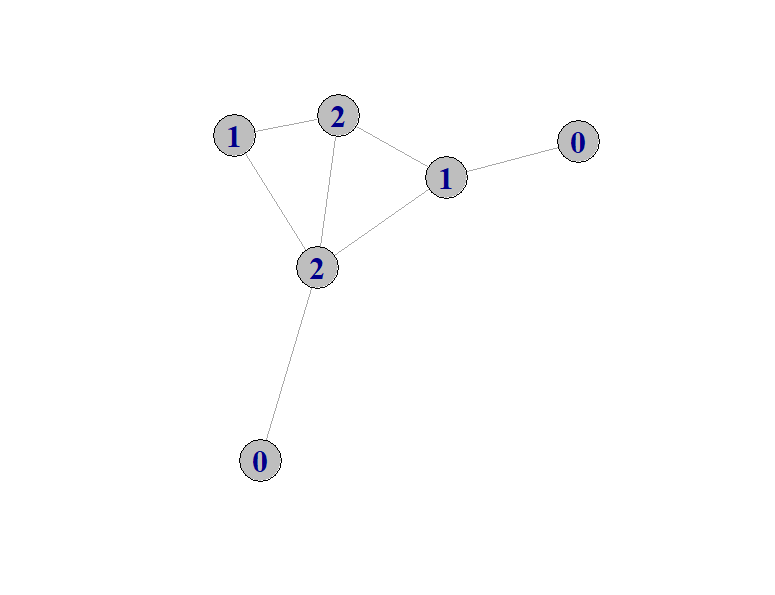}
\caption{Network with each node displaying Number of Triangles,  that node is involved with.\label{fig:No_Tri}}
\end{figure}

   \subsubsection{Number of triangles}
 From the adjacency matrix for our sample network Fig. \ref{fig:examgph}, we are able to calculate the number of triangles each node of the graph is involved in, which for nodes 1 to 6, we obtain
  0, 1, 2, 1, 2, 0  respectively. This is shown in Fig. \ref{fig:No_Tri}  where the number of the node represents the number of triangles that node is involved in forming.
  
  \section{Our Method}\label{sec:OurMeth}
\begin{figure}[!t]
\centering
\includegraphics[width=3.0in]{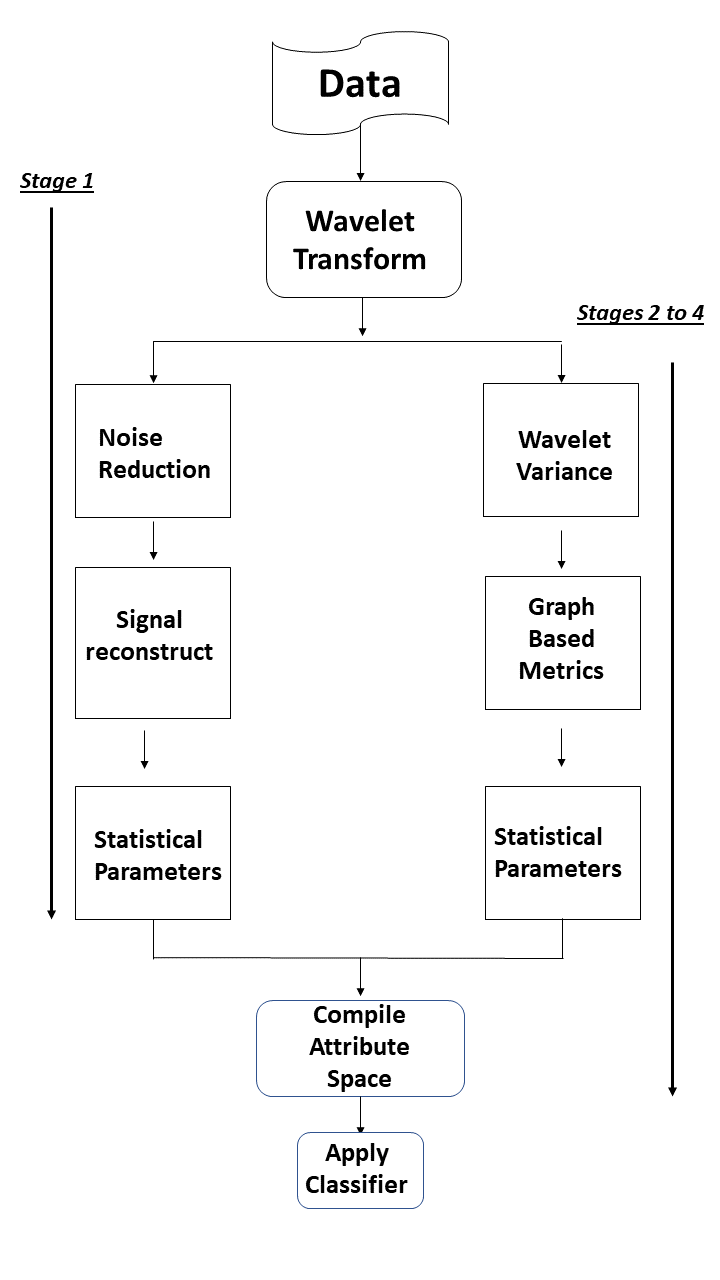}
\caption{Functional diagram of proposed method process. \label{fig:process1}}

\end{figure}

Our Four Stage methodology  relies heavily upon two separate properties of the wavelet transform, the first property being actual values of the coefficients at each wavelet decomposition scale, used in Stage 1.  The second property being the variance between the wavelet coefficients, at each scale, to determine correlation between signals, used in Stages 2 to 4, represented in a simplified block diagram, see Fig. \ref{fig:process1}.
We use the MODWT as this permits us to apply such only once to our data and utilise  the transformed data to reduce signal noise, derive connections between node/electrodes and implement various graph theoretic parameters to build the attribute space for the classifiers to be applied to. 

\subsection{Wavelet decomposition and synthesis:  Method Stage 1.}\label{sub:wavedecomp}
\begin{figure}[!t]
\centering
\includegraphics[width=3.0in, height = 2.4in]{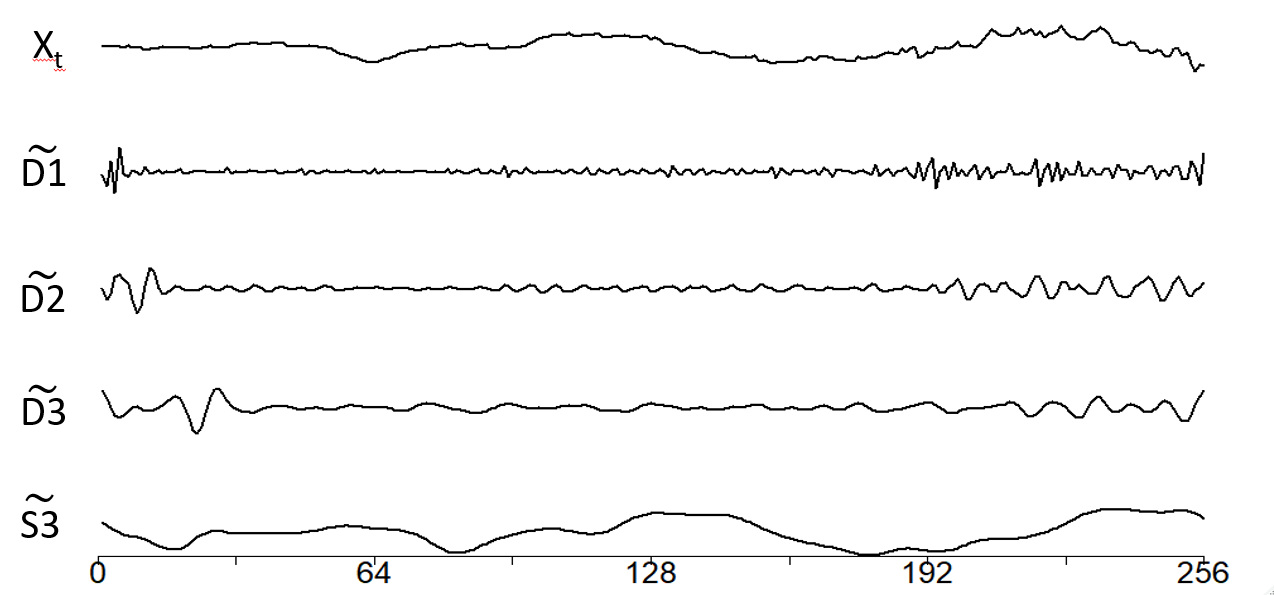}
\caption{$X_t$ Signal from an electrode for one epoch after Three level MODWT Wavelet decomposition into different bandwidths.\label{fig:WavedDComp}}

\end{figure}

By applying wavelets to our data we are able to decompose the time series $\{X_t\}$ from each electrode per epoch into nonoverlapping frequency bandwidths. These bandwidths are represented by the transform as the components within the vectors; $\mathcal{\widetilde{D}}_j \; \& \; \mathcal{\widetilde{S}}_J$, see Fig. \ref{fig:WavedDComp} as an example, with three level decomposition only. By utilising this decomposition we are able to select relevant frequency ranges that are related to the event to be studied.

After selection of the required frequency bandwidth we are able to reconstruct the signal using the inverse transform, also called synthesis\footnote{A reconstruction example is shown is Section \ref{subs:MethDCompE}. The inverse wavelet transform is fully described and documented \cite[Ch. 5]{P26Perc:2000}.} omitting the non-relevant frequencies, which would usually be contained within a subset of the different $\mathcal{\widetilde{D}}_j$ levels. This enables us  to derive statistical parameters and use such as attributes. i.e. mean, maximum and normalised energy\footnote{Normalised energy in $X_t$ defined as $\frac{\parallel X_t \parallel^2}{n} \equiv \frac{\langle X,X \rangle}{n} =  \frac{\sum_{t = 1}^{n} X^2_t}{n}$.
 } from this reconstructed signal.
 As an example of this see Table \ref{tab:ExAtt1i}. \par
 \textbf{Note:} An option is to use raw wavelet coefficients to derive the statistical parameters, however with $n$ coefficients for each relevant bandwidth, where we may need to derive statistical parameters from each level, \textit{(and for each electrode  \{1, \dots,K\})}.  This produces numerous potential attributes i.e. $K$ electrodes $\times\, J$ wavelet levels $\times$ Number of statistical parameters.

  \begin{table*}[ht!]
\centering
\caption{Data set resulting from wavelet decomposition 
 and wavelet synthesis  which omits unwanted frequencies}
\resizebox{.95\textwidth}{!}{%
 \begin{tabular}{||c| c c c c ||} 
 \hline
   & \multicolumn{4}{c||}{Attributes derived from each electrode's signal} \\
 $Individal_i$ & Elec-\textit{1} & Elec-\textit{2}  & ... & Elec-\textit{n}\\ 
 \hline\hline
 Epoch \texttt{1} &  $min_{11}, max_{11}, \ldots, normE_{11}$ &  $min_{12}, max_{12}, \ldots, normE_{12}$ & ...  & $min_{1n}, max_{1n}, \ldots, normE_{1n}$ \\ 
 Epoch \texttt{2}   &$min_{21}, max_{21}, \ldots, normE_{21}$ &  $min_{22}, max_{22}, \ldots, normE_{22}$ & ...  & $min_{2n}, max_{2n}, \ldots, normE_{2n}$\\
 \vdots & \vdots&   &  &\vdots \\
Epoch \texttt{m} & $min_{m1}, max_{m1}, \ldots, normE_{m1}$ &  $min_{m2}, max_{m2}, \ldots, normE_{m2}$ & ...  & $min_{mn}, max_{mn}, \ldots, normE_{mn}$ \\[1ex] 
 \hline
 \end{tabular}}
 \label{tab:ExAtt1i}
\end{table*}
 \subsection {Wavelet Correlation :  Method Stages 2 to 4.} \label{subs:WaveCorrel}
 The use of wavelets to estimate functional connectivity arises from previously noted results \cite{P22Acha:2007} and that wavelet based estimators of correlation 
 have desirable statistical properties \cite{P24Gutt:2000}. Here we are able to consider some topological properties  of networks or graphs that have shown some significance in relation to neural  and \textit{``small world"} networks  \cite{P29Lato: 2001}. First we are able to derive the nodes that are connected during an epoch and then develop indicators that alter highly between the seizure and non-seizure states. 
 \subsubsection{Number of Connections: Method Stage 2.}\label{subs:OMConn}
 Using the MODWT to decompose the time series from each electrode, per epoch we derive wavelet variance and similarly wavelet covariance together with  wavelet correlation between the transformed signals. Wavelet variance  is calculated at each level of wavelet decomposition  and is equal to the variance of the wavelet coefficients at that level. The sum of the wavelet variances from all decomposition levels equals the sample variance of the time series, similarly for wavelet covariance between each of the transformed signals \cite[Ch.~8]{P26Perc:2000}.
Covariance provides a measure of joint variation for two random sequences. The covariance of a random sequence with itself would be the variance, refer to Definition \ref{sam:covar}.  \par
 Hence, from wavelet variance as  outlined in Section \ref{subs:wavar},  we develop correlations by setting a user defined lower bound on the correlation value $\rho$, between two electrodes. If the correlation between the electrodes is greater than the lower bound  then we  define that a connection or an association, exists between these electrodes.  From these connections we may form an adjacency matrix, 
as an example see Table \ref{tab:ExAdl}.
From such an adjacency matrix it is possible to  construct a visualisation, which in this case is shown Fig. \ref{fig:examgph}, representing possible electrode connections during an epoch. The number of connections per node to other nodes may simply be observed as the sum of a row in the adjacency matrix. 
Using each degree of connection per node, (\textit{for each epoch}) we may form various indicators to use as attributes, i.e mean, maximum and skewness. 
We show an example of these attributes added to those in Stage 1, see Table \ref{tab:ExAtt2ai}.

\begin{table*}[t!]
\centering
\renewcommand{\arraystretch}{1.2}
\caption{Data set resulting from Stage 2, where connectivity statistics are added }
\resizebox{0.95\textwidth}{!}{%
 \begin{tabular}{||c| c c c c ||} 
 \hline
   & \multicolumn{4}{c||}{Attributes derived from Stage 1  and Stage2} \\
   & \multicolumn{3}{c}{Stage 1} & \multicolumn{1}{c||}{Stage 2}\\
 $Individal_i$ & Elec-\textit{1}& ...  & Elec-\textit{n}   & Connectivity \\ 
 \hline\hline
 Epoch \texttt{1} &  $min_{11}, max_{11}, \ldots, normE_{11}$ &...  &$min_{1n}, max_{1n}, \ldots, normE_{1n}$,  & $meanC_1, maxC_1, \ldots, skewC_1$ \\ 
 Epoch \texttt{2}   &$min_{21}, max_{21}, \ldots, normE_{21}$ &...  &$min_{2n}, max_{2n}, \ldots, normE_{2n}$, & $meanC_2, maxC_2, \ldots, skewC_2$\\
 \vdots & \vdots&   & \vdots  &\vdots \\
Epoch \texttt{m} & $min_{m1}, max_{m1}, \ldots, normE_{m1}$ &...  &$min_{mn}, max_{mn}, \ldots, normE_{mn}$,  & $meanC_m, maxC_m, \ldots, skewC_m$ \\[1ex] 
 \hline
 \end{tabular}}
 \label{tab:ExAtt2ai}
\end{table*}

\subsubsection{Global Efficiency: Method Stage 3.}
The efficiency of a network is a measure of how efficiently it exchanges information.
It has previously been applied to monitor brain complexity and interconnectivity \cite{P30Lato:2002}. For this indicator  we produce a vector containing the efficiency for each node of the graph, derived from the adjacency matrix constructed in Section \ref{subs:OMConn},   for each $\text{node}_i$ or $\text{electrode}_i$ in our network $G$ we compute  the reciprocal of the  shortest path length ($R_{i,j}$)  to each $\text{node}_j$ . For simplicity, it is usual to set the length of a single direct link to 1.
   For example in Fig. \ref{fig:examgph} the shortest path length from node 3 to node 6 is two, therefore $R_{3,6}$ is $\frac{1}{2}$. For each $\text{node}_i$  we then compute \newline $ (\sum_{j=1}^N R_{i,j}) / (N-1)$ where $N$ is the number of nodes and $R_{i,i} = 0$. Then for each epoch we calculate statistical parameters such as mean, max and skewness of the set of associated efficiency values.  We then add to our data set, as shown in Table \ref{tab:ExAtt3ai}. 

     \begin{table*}[t!]
\centering
\renewcommand{\arraystretch}{1.3}
\caption{Data set resulting from  Stage 3. where Global efficiency statistics are added}
\resizebox{0.95\textwidth}{!}{%
 \begin{tabular}{||c| c c c c ||} 
 \hline
   & \multicolumn{4}{c||}{Attributes derived by Stage 1, Stage 2 \& Stage 3} \\
  & Stage 1 &  & Stage 2 &  Stage 3 \\
 $Individal_i$ & Elec-\textit{1 to n}&   & Connectivity   & Global Efficiency \\ 
 \hline\hline
 Epoch \texttt{1} &  $min_{11}, max_{11}, \ldots, normE_{1n}$, &  &$meanC_1, maxC_1, \ldots, skewC_1$,  & $MeanG_1, MaxG_1, \dots SkewG_1$ \\ 
 Epoch \texttt{2}   &$min_{21}, max_{21}, \ldots, normE_{2n}$, &  &$meanC_2, maxC_2, \ldots, skewC_2$, & $MeanG_2, MaxG_2, \dots SkewG_2$\\
 \vdots & \vdots&   & \vdots  &\vdots \\
Epoch \texttt{m} & $min_{m1}, max_{m1}, \ldots, normE_{mn}$, &  & $meanC_m, maxC_m, \ldots, skewC_m$,  &  $MeanG_m, MaxG_m, \dots SkewG_m$ \\[1ex] 
 \hline
 \end{tabular}}
 \label{tab:ExAtt3ai}
\end{table*}

\subsubsection{ Eccentricity \& No. of Triangles:}\label{subs:eccen}
Another measure related to connectivity within small world networks works is eccentricity $\epsilon (i)$ of $node_i$, the greatest distance between $node_i$ and any other $node_j \in G$ : $\epsilon (i)=\max _{j\in G}d(i,j)$. Or how far a node is from the node most distant from it in the graph. 
 We ignore node pairs that are in different graph components and isolated vertices have eccentricity equal to zero.
 We would use the same adjacency matrix as used for Global efficiency.
 \noindent
 Apply that to our sample network, Fig. \ref{fig:examgph}  where each link has an assumed length of 1. From  nodes 1 to 6 we obtain \{3, 2, 2, 3, 2, 3\} respectively.
 From which we could derive statistical parameters to be used as attributes i.e. mean, maximum and sum.
 \paragraph{Number of triangles}
  Transitivity is the overall probability for the network to have adjacent nodes interconnected. A common way to quantify the transitivity of a network is by the means of the fraction of transitive triples \cite{P30Lato:2002}. It is calculated by the ratio between the observed number of closed triplets\footnote{A set of 3 points \textit{(or nodes)} with connections between them forming a triangle we define as a closed triplet or triple.} and the maximum possible number of closed triplets in the graph. 
  As we have the same number of nodes or electrodes in each epoch,  we drop the requirement of dividing  by the maximum possible number of formed triples in the epoch. 
 From the adjacency matrix for our sample network Fig. \ref{fig:examgph}, we are able to calculate the number of triangles each node of the graph is involved in which for nodes 1 to 6, we would obtain
  \{0, 1, 2, 1, 2, 0\}  respectively. We show this in Fig. \ref{fig:No_Tri}  where the number of the node represents the number of triangles ($No.Tri$) that node is involved in forming.
   We use this as an additional indicator,  and place with  the attributes derived in Section \ref{subs:eccen}, formed by each node per epoch. Similarly, we add this set of attributes to those formed previously, providing us with a final data set to which we attached class labels to each epoch, see Table  \ref{tab:ExAttalli}, which represents Step 4. We then apply a  classifier to this transformed data set.  As we may expect a large imbalance in our data regarding seizure compared to non-seizure records, we chose a cost sensitive decision forest algorithm to overcome such an imbalance. 
   \begin{table*}[ht]
\centering
\renewcommand{\arraystretch}{1.4}
\caption{Data set resulting from the 4 Steps after attaching graph based metrics  and class label}
\resizebox{0.95\textwidth}{!}{%
 \begin{tabular}{||c| c c c c c||} 
 \hline
   & \multicolumn{5}{c||}{Attributes derived from each Step } \\
 $Individual_i$ & \textit{Stage 1}  &\textit{Stage 2}   & \textit{Stage 3}&  \textit{Stage 4}& label \\ 
 \hline\hline
 Epoch \texttt{1} &  $min_{11},\ldots, normE_{1n}$,& $meanC_1,\ldots, skewC_1$, & $MeanG_1,\dots, SkewG_1$,& $HmeanE_1,\ldots, No.Tri_1$ & non-seizure\\ 
 Epoch \texttt{2}   &$min_{21},\ldots, normE_{2n}$,& $meanC_2,\ldots, skewC_2$, & $MeanG_2,\dots,SkewG_2$, &$HmeanE_2,\ldots, No.Tri_2$& seizure\\
 \vdots & \vdots& \vdots   &\vdots  &  \vdots &\vdots     \\
Epoch \texttt{m} &  $min_{m1},\ldots, normE_{mn}$,& $meanC_m,\ldots, skewC_m$,  &$MeanG_m,\dots, SkewG_m$, &$HmeanE_n,\ldots, No.Tri_n$  & non- seizure\\[1ex] 
 \hline
 \end{tabular}}
 \label{tab:ExAttalli}
\end{table*}

 \section{Implementation of our Technique and Experimental results}

     \subsection{Software}
 For this work  we used publicly available software;  WEKA \cite{P31Witt:2009} together with R \cite{P32Cran:2019},  including the R modules: wmtsa \cite{P33Cons:2018}, Brainwaver \cite{Acha:2012}, dplyr \cite{P35Wick:2022}, eegkit \cite{P2Helw:2018} and  igraph \cite{P36Csar:2006}.
  
   \subsection{Data}\label{subs:data}
 We accessed a  publicly available data set \cite{P38Gold:2000}, which was collected at the Children's Hospital Boston,
Massachusetts (MIT). This consists of EEG recordings  from paediatric subjects with intractable seizures. There are recordings from 24 different investigations from 23 subjects. The International 10-20 system of EEG electrode positions (\textit{for 23 electrodes}) was used for these recordings with the data sampled at 256Hz.  See https://archive.physionet.org/pn6/chbmit/ for further details and waveform views of this data . For this study we used a single file from each of 12 individuals, these individuals were identified in the MIT data by the case numbers; 1, 2, 3, 4, 5, 6, 7, 8, 10, 11, 23 and 24.
 The cases were selected on the basis of; the data across cases was reported  with same identical contiguous electrode sequence, i.e. $\{ FP1, F7, T7,\ldots, FT10, P8 \}$, each case  from a different individual, limited selection to files up to two hours in duration and no seizure event within the first 10 minutes. 
    Most of the selected files were approx one hour in duration. As the files have 256 data points per second for each electrode signal, therefore one hour equates to $3600 \times 256 = 921600$ data points, per electrode.  A few files were two hours in length,  for these files we selected either the first or second hour depending upon which hour had the first seizure event. This enabled us to have approx 1 hour of recording from each patient.
    The raw data had indicated periods of seizure, we subset the data into one second intervals i.e. an epoch length of one second, then attached labels indicating the class of that epoch, i.e. non-seizure or seizure.  For two of the patients there were no seizures  within the hour sample, we still included these two into our training sets but did not include testing upon them as all methods returned very high degree accuracy from these instances \textit{(as no seizures to detect)}.  Of these two patients, one simply had no seizures recorded in files of up to two hours duration, the other had the sequence in which the  electrodes were reported, altered before the first file with a seizure occurred. 
\begin{figure*}[!t]
\centering
\subfloat[During Non-Seizure]{\includegraphics[width=2.8in,height = 2.7in]{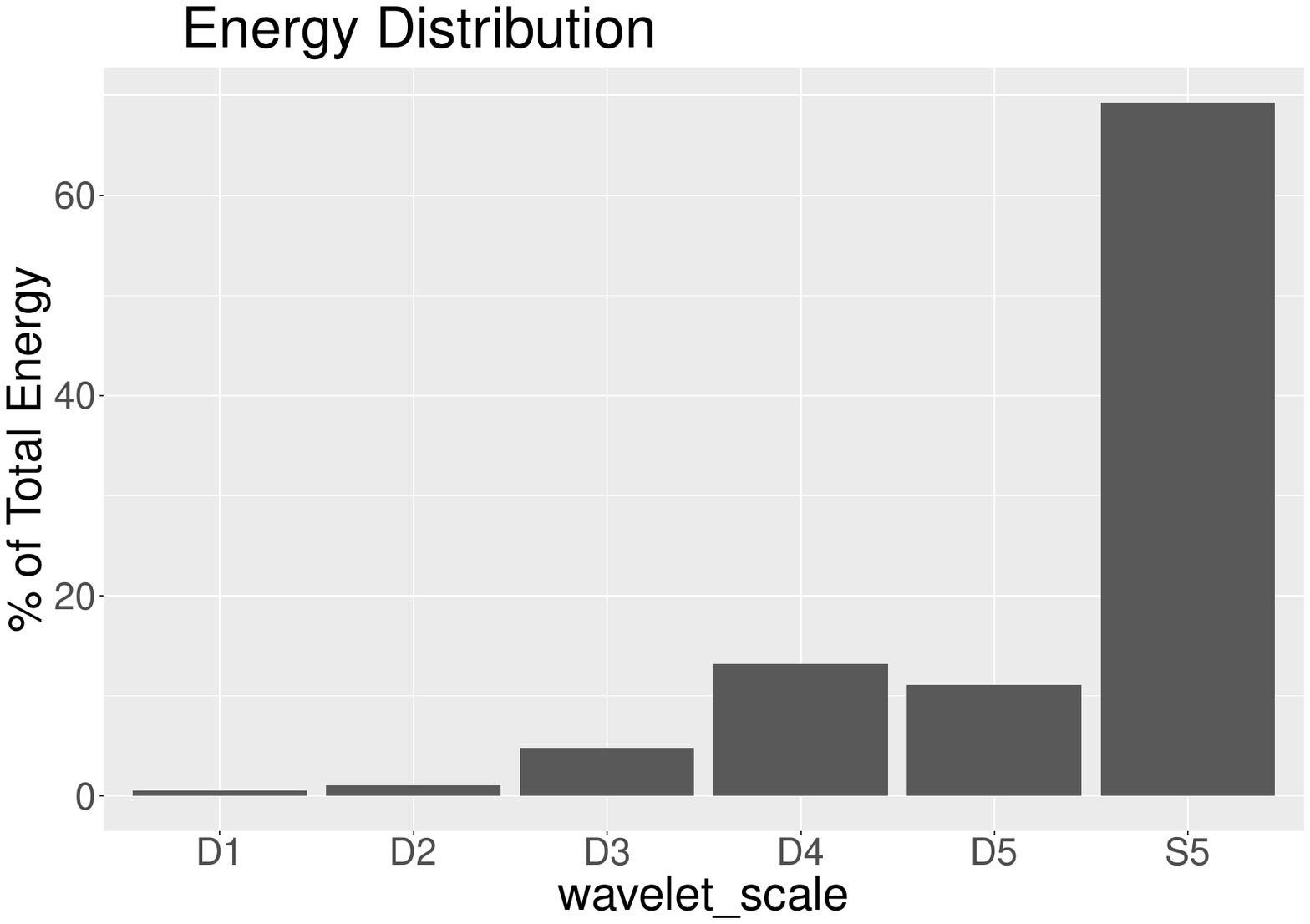}%
\label{fig:Att1a}}
\hfil
\subfloat[During Seizure]{\includegraphics[width=2.8in, height = 2.7in]{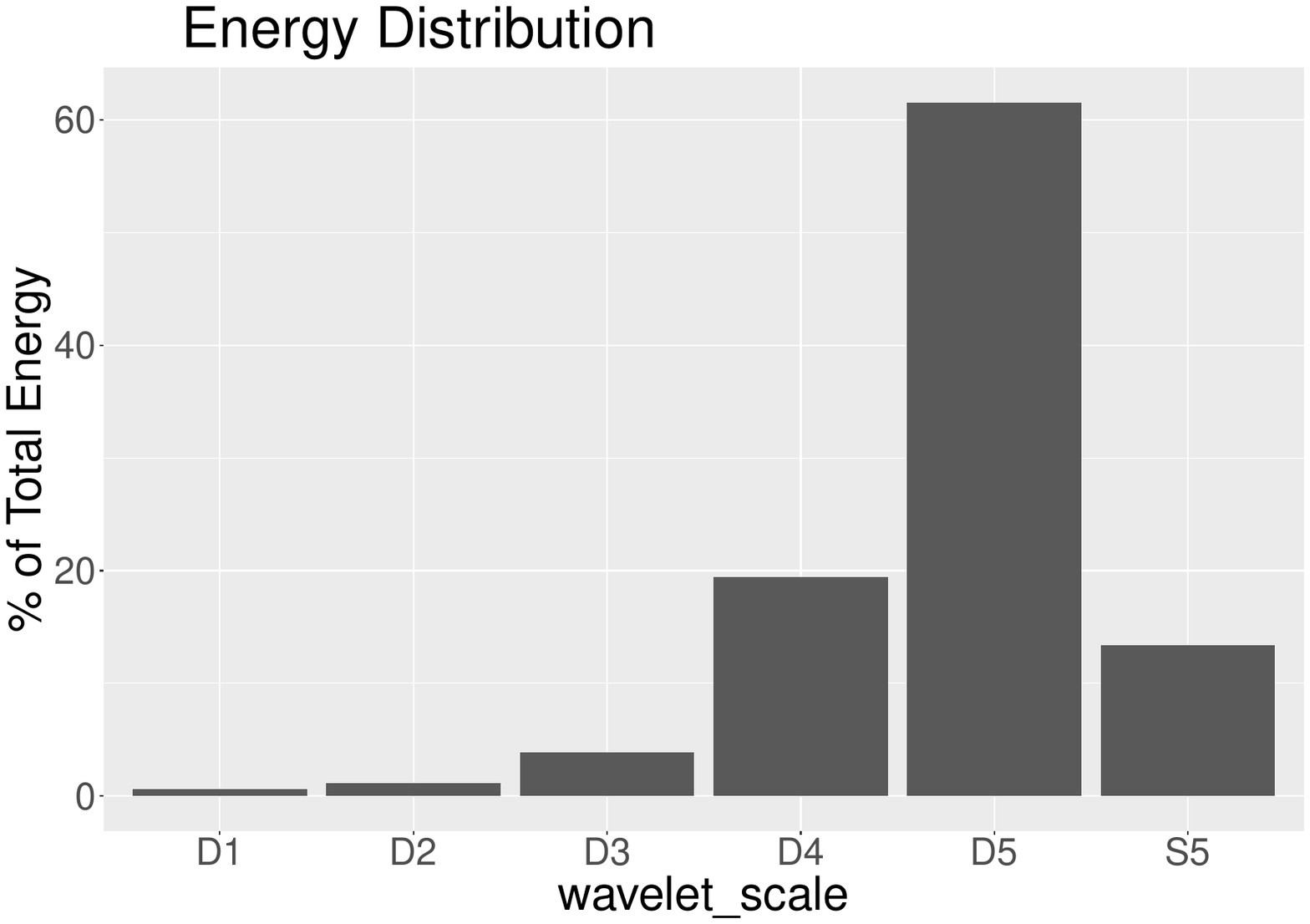}%
\label{fig:Att2}}

\caption{Energy distribution with Energy \% from an electrode during an epoch from each Wavelet level/frequency band  across the two Brain states.}
\label{fig:EngD}
\end{figure*}


        \subsection{Application of Our Method}%
    \subsubsection{Wavelet decomposition} \label{subs:MethDCompE}
\begin{figure*}[!t]
\centering
\subfloat[During Non-Seizure]{\includegraphics[width=2.8in, height = 2.8in]{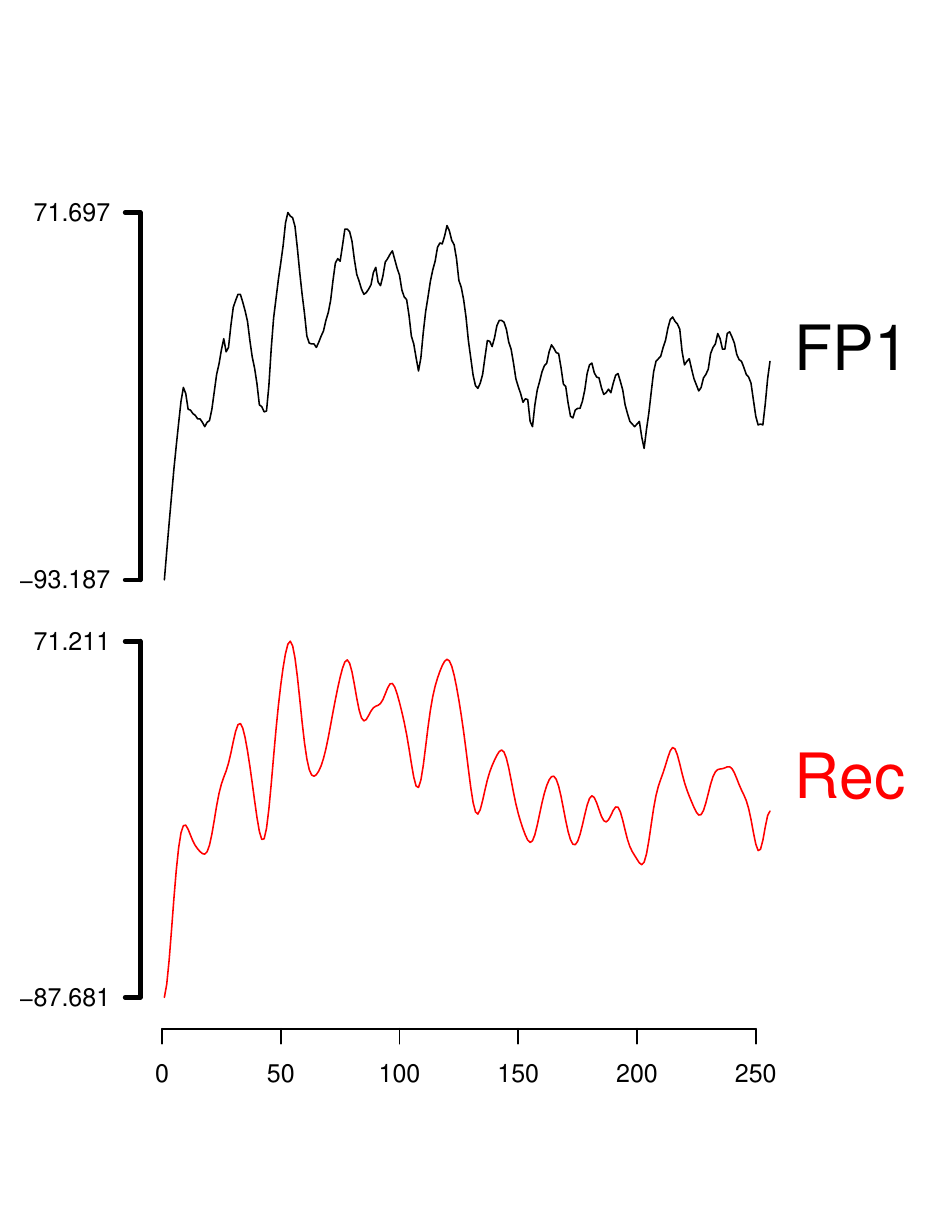}%
\label{fig:ResN}}
\hfil
\subfloat[During Seizure]{\includegraphics[width=2.8in, height = 2.8in]{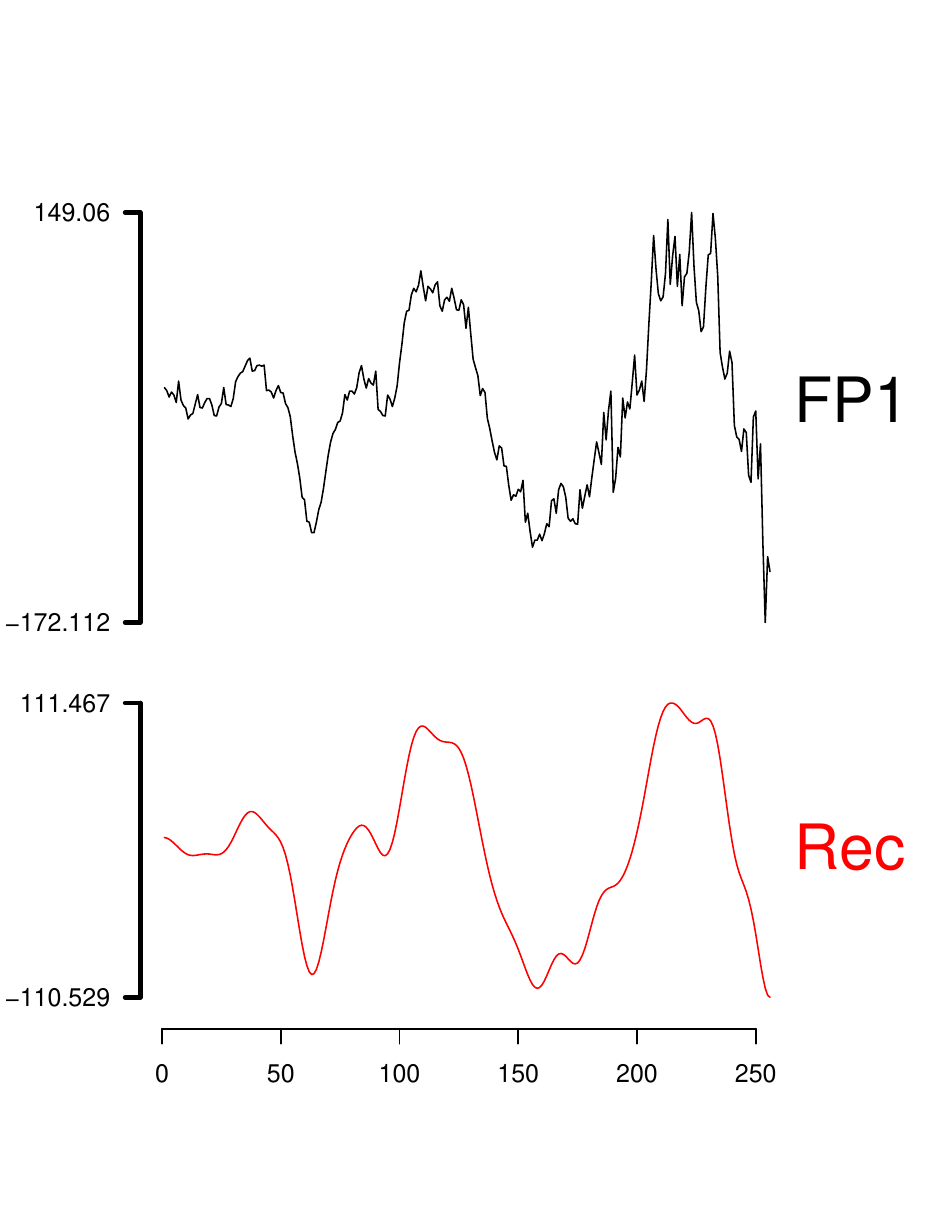}%
\label{fig:ResS}}

\caption{Microvolts during an epoch, for electrode FP1 and \underline{Rec}onstructed FP1 signal across the Brain States.}
\label{fig_Recon}
\end{figure*}

    The medically established  EEG frequency rhythms, given in Section \ref{sec:Intro} are shown in Table \ref{tab:freq}  mapped against their respective wavelet scale, these frequency bandwidth mappings result from the data being collected \textit{(or sampled)} at 256Hz. Hence, we used wavelet decomposition to 5 levels to achieve this frequency segmentation.  As noticed in other works regarding graphical properties of brain functional networks \cite{P22Acha:2007,Acha:2012} we used the wavelet, Least Asymmetric 8 (la8).
       As an example of the wavelet decomposition see Fig. \ref{fig:Att1a} which highlights  the energy distribution from an Individual electrode for an epoch  by the wavelet levels/bandwidths, during the non-seizure state. Compare this to Fig. \ref{fig:Att2}, easily observed are the changes in the Alpha, Theta and Delta rhythms  between the  seizure or non-seizure states. Also the first two wavelet decomposition levels representing the High Gamma and Gamma rhythms provide very little energy discrimination between either of the states . Therefore after application of MODWT we eliminate these two levels and reconstruct the signal using the inverse wavelet transform, to have a signal which is smoother and contains energy or information that is more relevant  i.e. less possibility of including noise, see Fig. \ref{fig:ResN} and Fig. \ref{fig:ResS}


         \begin{table}[ht!]
\caption{Frequency bandwidths resulting from  wavelet decomposition}
\label{tab:freq}
\centering
\resizebox{0.40\textwidth}{!}{%
 \begin{tabular}{||c| c c ||} 
 \hline
 Level & Frequency & Band  \\ [0.4ex] 
 \hline\hline
\raisebox{-0.5ex}[10pt]{$\widetilde{\mathcal{D}_1}$} & 64-128Hz & High Gamma  \\ [.5ex]
 \hline
 \raisebox{-0.5ex}[10pt]{$\widetilde{\mathcal{D}_2}$} & 32-64Hz & Gamma \\ [.5ex]
 \hline
 \raisebox{-0.5ex}[10pt]{$\widetilde{\mathcal{D}_3}$} & 16-32Hz & Beta  \\ [.5ex]
 \hline
  \raisebox{-0.5ex}[10pt]{$\widetilde{\mathcal{D}_4}$} & 8-16Hz & Alpha  \\ [.5ex]
 \hline
  \raisebox{-0.5ex}[10pt]{$\widetilde{\mathcal{D}_5}$} & 4-8Hz & Theta \\  [.5ex]
 \hline 
  \raisebox{-0.5ex}[10pt]{$\widetilde{\mathcal{S}_5}$} & 0-4Hz & Delta  \\ [0.5ex]
 \hline
\end{tabular}}
\end{table}

\subsubsection{Attribute building: } \label{subsec:Attrib1}

During this stage we used data from all 23 electrodes available. From the reconstructed signal for each electrode, per epoch we derive  statistical parameters as attributes. Here we used minimum, maximum, mean, standard deviation and normalised energy. We choose these statistical parameters as they are simple to implement, have been previously used \cite{P16Chen:2017, P15Jaco:2018}   and the ``Select Attribute" function within WEKA, provided high ranking scores in information gain for these attributes, across the various individuals. For the  signal from each electrode we place the attributes derived from each epoch in sequential time order, as an example see  Table \ref{tab:ExAtt1i}. \par
 With the aim improve performance  we  used wavelet variance to develop additional attributes based upon Graph theory, as mentioned in Section \ref{subsec:GrpT}. 
 \subsubsection{Connections}\label{subs:Conns}

Wavelet correlation  which is derived from wavelet covariance see (\ref{wave:corel}), provides  a metric allowing us to determine association or joint variability, between the transformed signals from each electrode to another, during an epoch. We set a data driven lower bound\footnote{Altering the value of the lower bound has a direct affect on the number of connections available. To be able to derive graphical based indicators and statistical parameters from the number of connections, the lower bound was adjusted to ensure we had at least a small average number of derived electrode connections per epoch, during the seizure state. In this situation $\rho = 0.125$.} on the correlation value $\rho$. If the correlation value $\rho$ between two transformed signals is greater than or equal to the defined lower bound we conclude that a connection exists. between the two during the epoch.

\begin{figure*}[!t]
\centering
\subfloat[During Non-Seizure]{\includegraphics[width=2.8in, height= 2.7in]{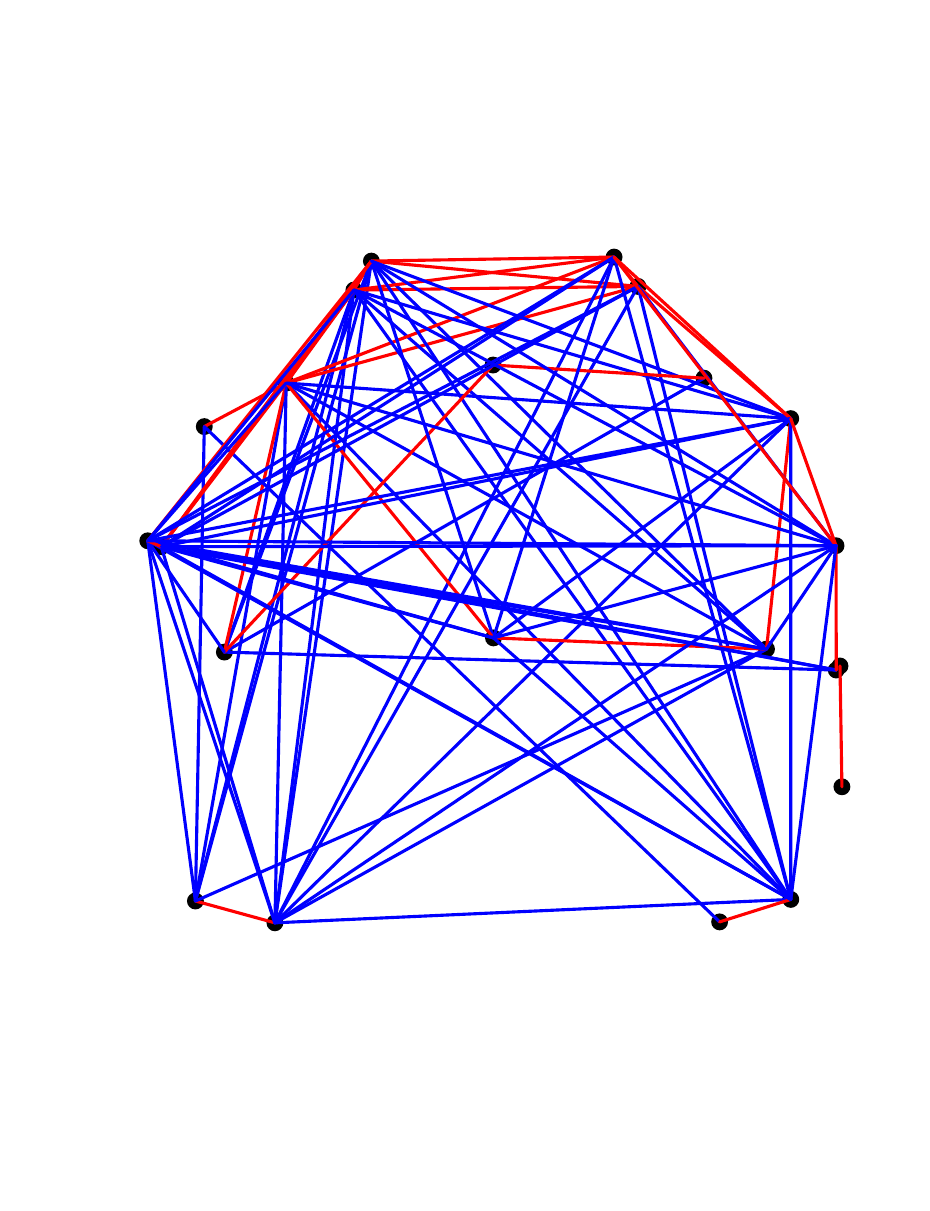}%
\label{fig:EEGConns}}
\hfil
\subfloat[During Seizure]{\includegraphics[width=2.8in, height = 2.7in]{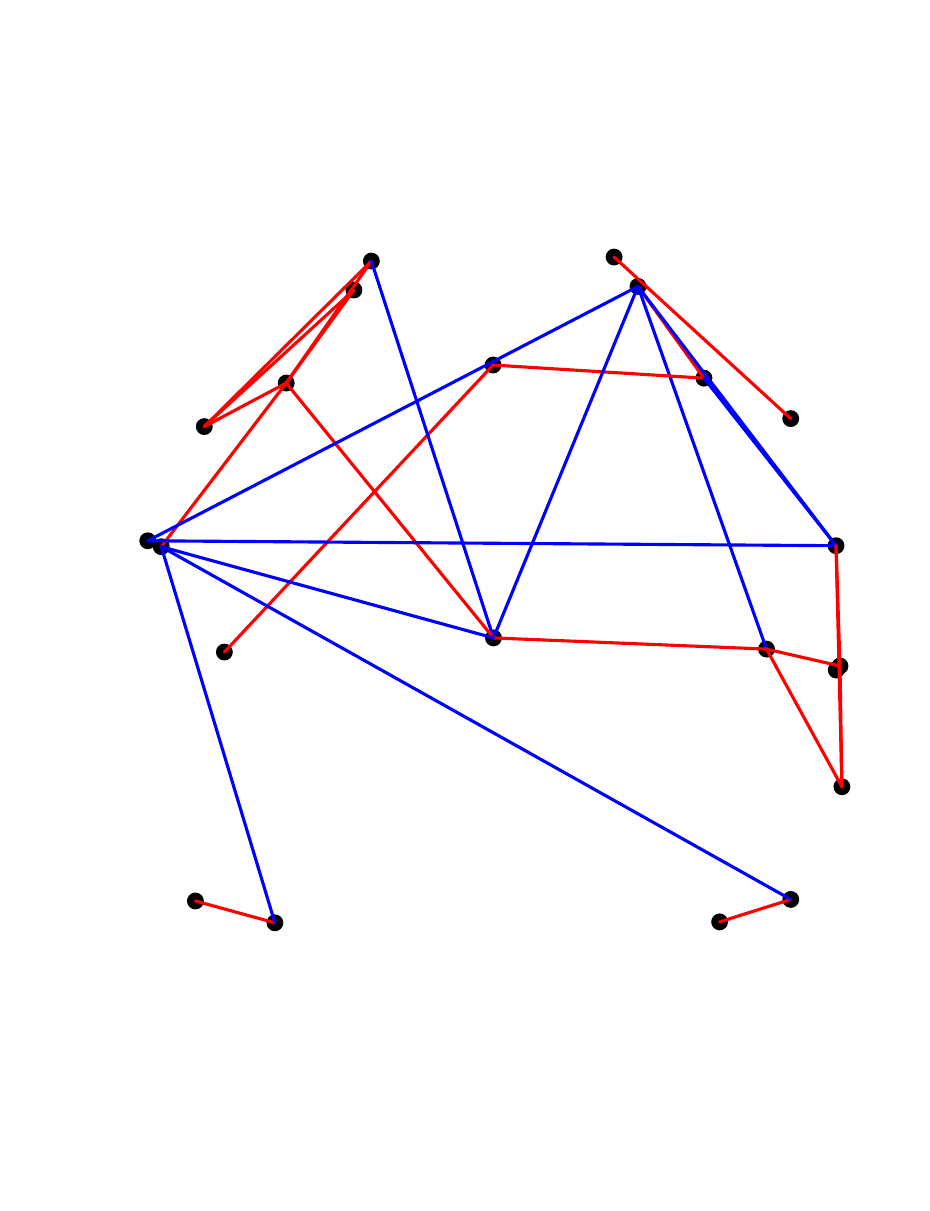}%
\label{fig:EEGConnsS}}

\caption{Derived EEG electrode connections, for an epoch, and a single wavelet level at a defined $\rho$ across the Brain States.}
\label{fig_ElecconK}
\end{figure*}

As an example resulting from our data, see Fig. \ref{fig:EEGConns} and Fig \ref{fig:EEGConnsS} which are plan representations of the derived connections on the EEG cap. 
The black dots represent the electrode positions. These representations are for one epoch at a single wavelet level and an arbitrarily chosen correlation lower bound. This highlights how the number of connections across electrodes alter during the different brain states. The 2D image representations of the 3D shape distorts distance. Therefore  the two colours simply represent actual length of a connection. Here red signifies a distance less than or equal to 10 cm, blue longer. 

From the data we found the greatest variance in connectivity, between the non-seizure and seizure states within the wavelet levels $\widetilde{\mathcal{D}_1}\;\text{and}\;\widetilde{\mathcal{D}_5}$. Only a small percentage of signal power  was observed at $\widetilde{\mathcal{D}_1}$, see Fig. \ref{fig:EngD}. It has been observed that variance within the Gamma frequency band may be used to determine difference within the brain states \cite{P39Alva:2014}. Similarly Theta band signals have shown an association with seizure \cite{P40Wanf:2022}. Hence we used the two levels, 
$\widetilde{\mathcal{D}_1}\;\text{and}\;\widetilde{\mathcal{D}_5}$  to derive connections and construct an  adjacency matrix, where a connection is represented by the value 1. In this situation we actually constructed two adjacency matrices, one for each chosen frequency bandwidth. As both matrices are the same size, we took the sum of the two matrices and then forced any elements in the resulting matrix with a sum greater than or equal to  1 to be 1.

\subsubsection{Attribute building: Stage 2} From Stage 2 and onwards we only used the data from 22 electrodes as one electrode had a high negative correlation to another electrode, and was omitted. From an adjacency matrix constructed using each electrode 
 in our network, we are able to  determine a degree of connection or number of direct connections during the epoch. 


From these row sums, representing  degree of connection,  we derive the statistical parameters: mean, maximum, 1st Quartile, 3rd Quartile and skewness. This provides us with additional attributes for the  epoch in question. We choose these statistical indicators from a varied range of easily calculated parameters and using WEKA determined the most suitable regarding information gain across the individuals' training sets. These additional attributes for each epoch are combined  with the data set outlined in Section \ref{subsec:Attrib1} increasing its dimension. An example of this updated data set is shown in  Table \ref{tab:ExAtt2ai}.

\subsubsection{Global Efficiency}\label{sub:eleceff}
 Section \ref{subs:Conns} highlights that given wavelet coefficients  we have a  methodology for the  construction of an adjacency matrix derived from the correlation between the electrodes.  From an adjacency matrix we are able to construct a visual representation of the possible network or Graph. Fig \ref{fig:examgph} shows a graph that may be derived from   Table \ref{tab:ExAdl}.  In this Graph  an electrode is represented as a numbered node and a pair of electrodes/nodes having a correlation above the threshold  are connected through an edge between them.
This graph has six nodes/vertices i.e. N = 6, and seven edges. In our previous work \cite{23Gran:2021} each single edge had an assumed unit length of 1, however it was later found  that by weighting these edges by an approximation of their length, slightly better discrimination permitted  a reduction in false negatives,  i.e.  Macro Average Recall improved from 0.64 to 0.72. For these distances across all individuals we used the EEG electrode spatial coordinates $\{x_i,y_i,z_i\}$ given in the R module eegkit \cite{P2Helw:2018}.  From a constructed  network diagram,  (\textit{and hence the adjacency matrix}) we derive how “efficiently” the electrodes are communicating across their network of connections. We calculate the global efficiency ($Eglob_i$) for each $\text{node}_i$. 

As an example using Fig. \ref{fig:examgph}, for  every $\text{node}_i$ we then compute  the reciprocal of the shortest path length ($R_{i,j}$)  to each other $\text{node}_j$, where 0 = the path length for an isolated node.  If we let the distance of each link = 1 then the shortest path length from node 1 to node 2 is one, hence $R_{1,2}$ is $\frac{1}{1}= 1$. Similarly, for the other nodes path connected  to node 1. i.e. $R_{1,3} = 0.5$.  For each $\text{node}_i$ we then compute  
 $Eglob_i =(\sum_{j=1}^N R_{i,j})/(N-1)$ where $N$ is the number of nodes in the entire network and $R_{i,i} = 0$.


 As we have a set of the spatial coordinates for the nodes, or electrodes under the ``10-20" system,  we are able to calculate the euclidean distance between the electrodes i.e. $d = (x^2 + y^2 + z^2)^{\frac{1}{2}}$,  and use these distances as the length of the link between each node, i.e. If there exists a direct connection between T9 to T10  (from left to right across scalp) instead of using 1 as the distance between the electrodes we used 15.67 which was derived from our chosen spatial coordinates set. We used these spatial coordinates for all such link calculations of then  and derived the $Eglob$ for each electrode per epoch.
 
 \subsubsection{Attribute building: Stage 3}
 Global efficiency for any $\text{node}_i$ during an epoch varies as  $0 \le Eglob_i \le 1$, as we are summing a maximum of N values, each the reciprocal of the distance between the electrodes and dividing the sum by N(N-1), where the lengths in our situation (using the eegkit coordinates) varied from 3.11 to 31.12 for a single link between directly connected electrodes. Hence, reciprocal of any sum of such values must be less than one. An isolated node/electrode has a defined efficiency value of 0.

From global efficiency values for an epoch we derive the statistical parameters: mean, maximum, 1st Quartile, 3rd Quartile \&  skewness.  These attributes are added to the data set, as an example see Table \ref{tab:ExAtt3ai}. Again the decision on which actual statistical parameters to use was guided by information gain provided via WEKA, across the individuals' training sets.
  \subsubsection{Eccentricity \& Number of Triangles}
 By utilising the same adjacency matrix as developed in Section \ref{subs:Conns}  it is possible to derive the eccentricity as described in Section \ref{subs:eccen}, of each electrode in the network and similarly the number of triangles each electrode is involved in forming, both across each epoch. These values are readily calculated using igraph \cite{P36Csar:2006} applied to the constructed Adjacency matrix.

 \subsubsection {Attribute building: Stage 4} As these parameters  are returned for each electrode we derive the following statistical parameters per epoch; harmonic mean, standard deviation and sum of eccentricity together with the sum of the number of triangles. Again these statistical indicators have been chosen as relevant by WEKA regarding information gain.  These attributes are then added to our data set,  as well as the classifying label that was attached when segmenting our original data into 1 second epochs. as an example see Table \ref{tab:ExAttalli}.
     Our derived data set resulting after the addition of Stage four construction provides us with only $5 \times$ 23 + 6 + 6 + 4 = 131 entries (attributes) per row plus class label.
     \vspace{-0.5cm}
\subsection{ Synopsis of  results and comparisons } \label{sub:results}
 We are mainly concerned with detecting a True Positive (TP) which refers to events  where epochs labelled  seizure are classified as seizure. Similarly, False Negative  (FN) refers to events where seizure labelled epochs were classed as non-seizure.
True negative (TN) refers to epochs labelled non-seizure and classified non-seizure and False Positive (FP) refers to an epoch labelled non-seizure and classified seizure.
 We define  following metrics or indicators:
 \begin{definition}
 \vspace{-0.4cm}
\begin{align*}
Accuracy\,\% & =  \frac{TP + TN}{TP + FN + TN + FP } \times 100\\
Recall  & =  \frac{TP}{TP + FN} \;  \hspace{5em} \text{see \cite{P37Powe:2011}}\\
Precision & =  \frac{TP}{TP+FP}\\
f-score & = \frac{2}{(Recall)^{-1} + (Precision)^{-1}}
\end{align*}
\end{definition}
We use  these indicators to determine effectiveness across the individuals in the study. As we are interested in correctly classifying seizure which is only a small proportion of the available data records, we  use a cost sensitive/class imbalanced algorithm. Here we choose CSForest \cite{P41Sier:2015} and applied it to our transformed data, shown in Table \ref{tab:ExAttalli}. The cost penalty structure for the algorithm is determined on a per individual basis from the relevant individual's Training data set. We used the leave-one-out cross validation method across all individuals, therefore an individual's training set comprised of all the other eleven individuals in the study.\par
We compared our results to five other methods as  outlined in Section \ref{sub:CopM}. Three of the comparative methods \cite{P16Chen:2017, P41aSrestha:2019, P42Zia:2021} had used the same data set in their development  that we used to test our method, see Section \ref{subs:data}.  We applied all the comparative methods, using each respective model's initial parameters,  to our data set  which was segmented into 1 second epochs and  used the  leave-one-out cross validation for all methods. 
For this comparison we used the macro average (\textbf{MA}) method, see Definition \ref{def:maf} where $R_i$ \& $P_i$ represents the Recall and Precision values for individual $i$.

\begin{definition} \label{def:maf}
\begin{align*}
\text{MA Recall (MAR)} & = \frac{R_1 + R_2 + \dots + R_N}{N}\\
\text{MA Precision (MAP)} & = \frac{P_1 + P_2 + \dots + P_N}{N}  \qquad \text{see \cite{P43Toro:2014}}\\
\text{MA f-score} & =  \frac{2}{\frac{1}{MAR}+\frac{1}{MAP}} \\
\end{align*}

\end{definition}

The overall results are shown in Table \ref{table:fscore}.
\begin{table*}[t!]
\caption{Macro Average results via comparative method}
\centering
\label{table:fscore}
\resizebox{.60\textwidth}{!}{%
 \begin{tabular}{||c c c c c||} 
 \hline
 Method No. & Method origin  &  f-score &  Recall & Precision  \\ [.2ex] 
 \hline\hline
 \rule{0pt}{12pt}{1} & Siddiqui et al. \cite{P14Sidd:2018}&0.42  &  0.35 & 0.52\\ [.2ex]
 \hline 
 \rule{0pt}{12pt}{2} & Chen et al. \cite{P16Chen:2017} & 0.36  &  0.24 & 0.69\\ [0.2ex] 
 \hline
  \rule{0pt}{12pt}{3} & Elsa Jacob et al. \cite{P15Jaco:2018} &0.13 &  0.08& 0.34\\ [.2ex]
 \hline 
 \rule{0pt}{12pt}{4} & Shrestha et al. \cite{P41aSrestha:2019}& 0.27 &  0.18& 0.55\\ [0.2ex] 
 \hline
  \rule{0pt}{12pt}{5} & Zia et al.\cite{P42Zia:2021} & 0.33 &  0.37& 0.39\\ [0.2ex] 
 \hline
  \rule{0pt}{12pt}\textbf{6} & Our Proposed Technique & \textbf{0.51}  & 0.72 & 0.39 \\ [0.2ex]
 
 \hline

 \hline
\end{tabular}}
\end{table*}

\subsubsection{ Metrics chosen }\label{metschos}

 The aim here is to detect all seizures and classify them as  seizures, \textit{(True Positives)}. Classifying non-seizures as seizures, \textit{(False Positives)} is at worst a waste of resources however, classifying seizures as non-seizures, \textit{(False Negatives)} may lead to catastrophic outcomes.  Hence, we consider that False Negatives should be minimised and have concentrated on development of attaining a high Recall result. 

 Here Recall \textit{(also known as Sensitivity)} is a metric for the proportion of seizures identified correctly, while Precision is  the proportion of seizure identifications that are actually correct.
   The f-score weights both Recall and Precision equally, our methodology  $\mathbf{ Method \;No \; 6\,}$  in both Tables \ref{table:fscore} \& \ref{table:comp} provides  a higher f-score and Recall values at a Macro average and for most at the individual level. While results for one individual \textit{Ind No.} 10 our method was not the optimal,  some of the comparative methods failed completely on specific individuals regarding these indicators. 
    It has been noted that many detection systems have imperfect
predictive value with a substantial, but variable number of false positive detections \cite{P1Tatum}, which may be noticed in the  wide range of Precision values across the methods/individuals.
 
The high class imbalance in the data renders using the AUC or area under the ROC curve as an unsuitable metric \cite{P44Sait:2015}, this imbalance still allowed a high Accuracy \% to be returned across all methods. As an example, in Table \ref{table:comp}, \textit{Ind No.} 3 with 3600 records, 48 were labelled seizure. If a classifier simply classed all records as non-seizure then an Accuracy of 98.67\% would result. Simply using Accuracy \% to evaluate performance may lead to erroneous results. Hence for the two individuals that did not have a recorded seizure event, as mentioned in section \ref{subs:data}, they have been omitted from Table \ref{table:comp}.

\begin{table*}[ht]
\caption{Statistical results via Individual across all comparative methods, using One second Epochs}
\label{table:comp}
\begin{center}
\resizebox{.95\textwidth}{5cm}{%
\begin{tabular}{ |c|c||c|c|c|c||c|c|c|c|c|c|  }
 \hline
 \multicolumn{12}{|c|}{\bfseries Comparative results} \\
\multicolumn{12}{|c|} {\hspace{0.5cm}\textbf{Metric}\hspace{11cm}\textbf{Metric}} \\
 \hline
 \textit{Ind.No.}& Method & \textbf{Accuracy\%} & \textbf{Recall} & \textbf{Precision} &\textit{f-score}& \textit{Ind.No.}  & Method & \textbf{Accuracy\%} &\textbf{Recall} & \textbf{Precision}& \textit{f-score}\\
 \hline
  1 & 1  & 98.94    & 0.073&   1.00 & 0.136 & 5& 1 &97.47& 0.776 &0.581 & 0.664 \\
      &2 &   98.87   & 0.098   &1.00&  0.178   & &  2  &98.00  &0.379& 1.00& 0.550\\
      &3 &  97.69        &0.00&  0.00&   0.00      & &3 & 96.89  &  0.043& 0.833& 0.082\\
      & 4  &99.14 & 0.248   &   1.00 &  0.392 &    &4 & 98.22& 0.509 & 0.894& 0.648 \\
       & 5  &97.50 & 0.024   &   0.020 &  0.0.22 &    &5 & 96.64& 0.819 & 0.487& 0.611 \\
     & $\mathbf{6}$  &$\mathbf{99.61}$ & $\mathbf{0.850}$    &  $\mathbf{0.810}$ &  $\mathbf{0.829}$  &   &$\mathbf{6}$ & $\mathbf{98.22}$ & $\mathbf{0.784}$  &$\mathbf{0.679}$  &  $\mathbf{0.728}$ \\
    \hline
 2 &1  & 74.77    &0.488&  0.044 &0.081 & 7 & 1  &96.48& 0.330 &0.327 & 0.328\\
    &2&   97.46 & 0.037  &0.200 &  0.062  & &  2  & 98.87   &0.567 & 1.00 &0.724\\
    &3 &97.72 & 0.012&  0.500   &  0.024  &      & 3 & 98.61&  0.569 & 1.00&0.725\\
   & 4   &97.85 & 0.280& 0.561&  0.374&  &4 & 98.76& 0.576 &1.00 &0.689\\
     & 5   &97.05 & 0.402& 0.367&  0.384&  &5 & 96.48& 0.680&0.398 &0.502\\
   & $\mathbf{6}$   &$\mathbf{99.50}$  & $\mathbf{0.720}$ &$\mathbf{ 0.298}$ & $\mathbf{0.421}$  &  &$\mathbf{6}$  & $\mathbf{94.90}$ &$\mathbf{0.866}$ &$\mathbf{0.323}$ & $\mathbf{0.471}$\\
   \hline
 3 &1& 97.94    &.208&   0.217& 0.213 & 8 & 1  &95.39& 0.241 &0.476 &0.320\\
    &2&  98.72  & 0.021   &1.00& 0.042& &   2 & 95.61   &0.018 &1.00 &0.037\\
    &3&98.67 & 0.00&  0.00&   0.00         &  & 3 &95.50 &  0.00 &0.00 &0.00\\
   & 4 &98.72 & 0.042& 1.00 & 0.800 &   &4   & 95.55 & 0.105&0.531 &0.175    \\
     & 5 &97.61 & 0.229& 0.183 & 0.204 &   &5   & 95.28 & 0.272&0.458 &0.341    \\
   & $\mathbf{6}$   &$\mathbf{98.22}$&$\mathbf{ 0.784}$& $\mathbf{0.679 }$& $\mathbf{0.728 }$&   &$\mathbf{6}$    & $\mathbf{93.30}$ &$\mathbf{ 0.741}$&$\mathbf{ 0.374}$& $\mathbf{0.496}$\\
   \hline
     4 &1 & 98.67    &0.020&  0.500 & 0.039 & 10 & 1   &97.77& 0.820 & 0.420& 0.556\\
      &2&  98.36  & 0.00   &0.00 &0.00 &          &   2 & 99.61  &0.770& 1.00& 0.870\\
      &3 &98.67 & 0.00&  0.00& 0.00  &            &3   & 98.42 & 0.109 & 1.00&0.197\\
      & 4  &98.09 & 0.00 &  0.00 & 0.00 &    &4 & 98.94 & 0.875& 0.651&0.747\\
       & 5  &98.53 & 0.00 &  0.00 & 0.00 &    &5 & 97.97 & 0.734& 0.456&0.562\\
       & $\mathbf{6}$   &$\mathbf{88.56 }$&$\mathbf{ 0.347}$& $\mathbf{0.042}$ &$\mathbf{ 0.075}$ &    &$\mathbf{6}$  & $\mathbf{95.60 }$&  $\mathbf{0.875}$&$\mathbf{ 0.272}$& $\mathbf{0.415}$\\
     \hline
   23 &1 & 96.89    &0.009&   1.00 & 0.18 & 24 & 1   &98.91& 0.500 & 0.641&0.562\\
      &2&  96.86 & 0.00  &0.00& 0.00&  &        2  & 98.97   & 0.520& 0.667& 0.584\\
      &3 &95.19 & 0.001&  0.016& 0.11 &  &      3 & 95.56& 0.040 & 0.018& 0.024\\
      & 4   &96.86 & 0.00&  0.00 & 0.00 & &4 & 98.92 & 0.580 & 0.617 & 0.598\\
       & 5   &96.61 & 0.009&  0.091 & 0.016 & &5 & 98.28 & 0.580 & 0.414 & 0.483\\
      & $\mathbf{6}$   &$\mathbf{92.30 }$&$\mathbf{0.788}$&$\mathbf{  0.261}$ &$\mathbf{0.392 }$ & &$\mathbf{6}$  & $\mathbf{99.86 }$& $\mathbf{0.600}$ &$\mathbf{ 0.588}$ & $\mathbf{ 0.594}$ \\
  
 \hline
\end{tabular}}
\end{center}
\end{table*}

\begin{table*}[ht!]
\caption{Macro Average results via Stages within Our Proposed Technique}
\label{table:Multifscore}
\begin{center}
\resizebox{.60\textwidth}{!}{%
 \begin{tabular}{|c c c ||c c c |} 
 \hline
 \multicolumn{6}{|c|} {\hspace{0.5cm}\textbf{Stage 1 }\hspace{2.8 cm}\textbf{ Stages 2 to 4}} \\
  f-score  &   Recall &  Precision &  f-score  &   Recall &  Precision  \\ [.2ex] 
 \hline\hline
0.485 & 0.477 & 0.493  &  0.10 & 0.215 & 0.066\\ [.2ex]
 
 \hline

 \hline

\end{tabular}}
\end{center}
\end{table*}

\begin{table*}[ht!]
\begin{center}
\caption{Statistical results via selected Individuals across Stages, within Our Proposed Technique}
\label{table:subIndW}
\resizebox{0.9\textwidth}{!}{%
\begin{tabular}{ |c||c|c|c|c||c|c|c|c|c|  }
 \hline
 \multicolumn{10}{|c|}{\bfseries Comparative results across Stages of Our Method } \\
\multicolumn{10}{|c|} {\hspace{0.5cm}\textbf{Metric  : Stage 1 }\hspace{6.5cm}\textbf{Metric :  Stages 2 to 4}} \\
 \hline
  Ind No. & \textbf{Accuracy\%} & \textbf{Recall} & \textbf{Precision} &\textit{f-score}&  Ind No. & \textbf{Accuracy\%} &\textbf{Recall} & \textbf{Precision}& \textit{f-score}\\
 \hline
   1  & 99.14    & 0.256& 0.883 & 0.392 &  1 &98.50& 0.077 &0.143 & 0.100 \\
      2 &   93.68   & 0.598  &0.202&  0.302  &  2  &92.96  &0.232& 0.091& 0.131\\
      10 &  97.02        &0.922&  0.219&  0.354     & 10 & 77.68  &  0.344& 0.028& 0.052\\
      24 &  98.92        &0.560&  0.662&  0.589      & 24 & 82.42 & 0 .380& 0.031& 0.057\\
 \hline
\end{tabular}}
\end{center}
\end{table*}

We may apply each of the two main parts of our method, which were mentioned in Section \ref{sec:OurMeth}, totally separately to gauge the contribution of these Stages. Table \ref{table:Multifscore}  demonstrates that Stage 1 with 115 attributes, returns only slightly better results in respect to the  comparative methods shown in Table \ref{table:fscore} and  the  Stages 2 to 4 (\textit{16 attributes}) provide only minor value for a few additional attributes. Similarly, this is further shown in Table \ref{table:subIndW} for 4 selected individuals. Hence, Stages 2 to 4 may not be  worthwhile by themselves for such a  limited number of  electrodes, however they provide  additional gain enabling our method to outperform the comparative methods, over the shorter one second epochs.
 
 Our methodology relies upon two separate properties of the wavelet transform. The first being  actual values of the coefficients at each wavelet decomposition scale, which was used in Our Method  Stage 1.  The second property being the variance between the wavelet coefficients to determine correlation between signals at specific frequency bandwidths, used in Stages 2 to 4. 
\subsection{Comparative methods}\label{sub:CopM}  We compared our method to five other published methods, we mirrored the methodologies in each method, using R and WEKA. Then applied these methodologies to the same raw data \textit{(using one second epochs, to approximate real time monitoring)} that we had used, defined in Section \ref{subs:data}. 
  The five works we compared to are as follows:
  
  \begin{itemize}
  \item [\textbf{1.}] Siddiqui, Islam and Kabir \cite{P14Sidd:2018} used nine statistical parameters, i.e. (Min, Max, \dots, Skewness)  derived from each electrode signal to build the attribute space, resulting in $ 9 \times 23  =  207$ attributes  plus label per record, then applied the forest classifier SysFor \cite{P45Isam:2012}, 100 trees. 
  \item [\textbf{2.}] Chen, Wan, Xiang and Bao \cite{P16Chen:2017}, used the DWT   to decompose the EEG signal into different frequency bands, selected a subset of these frequency bands and  derived statistical parameters from the wavelet coefficients in the chosen frequency bands (wavelet decomposition scales) then applied the classifier SVM, (via a leave one out methodology), to their transformed data.  This was  repeated for different mother wavelets,  (i.e. Daubechies, Haar,  Symlets, etc ...).
  For comparison, we used this publication's methodology for the \textit{Haar} wavelet, being six frequency bands together with three statistical parameters: Minimum, Standard deviation and Skewness. This resulted in $6 \times 3 \times 23 = 414$ attributes plus label  per record. 
  \item[\textbf{3.}] Elsa Jacob, Nair, Iype and Cherian \cite{P15Jaco:2018}  initially smoothed the signal, using Total variation denoising, (\textit{also known as total variation regularisation}), then applied the DWT using the Daubechies 4  wavelet (db4) to extract relative energies at 4 different sub-bands. This provides $ 4 \times 23  = 92$ attributes plus label. An SVM classifier was then applied to this data.
   \item[\textbf{4.}] Shrestha, Dahi Shrestha and Thapa \cite{P41aSrestha:2019} used the DWT, with a Daubechies 8 wavelet (db8)  to extract 5 sub-bands and used relative energies from three of these sub-bands.This provided $3 \times 23 =  69$ attributes plus label. Then applied an ANN with backward propagation using a hidden layer of size 50, to this overlapped and transformed data.  
   
     \item[\textbf{5.}] Zia, Qureshi, Afzaal,  Qureshi, and Fayaz \cite{P42Zia:2021} used the DWT, with a Daubechies 4 wavelet (db4)  to extract 5 sub-bands  and generate 6 statistical parameters from each sub-band per electrode.This provided $5 \times 6  \times 23 =  690$ attributes plus label. Then used a KNN-1 classifier to this overlapped and transformed data. 
  
\end{itemize} 
 \section{Conclusion}
 In this study we have combined two  inherent capabilities of the MODWT to  decompose the signal into discrete bandwidths and  wavelet variance to develop frequency cross-correlation between signals within selected bandwidths. 
  This was achieved by using the la8 wavelet and 5 level wavelet decomposition. 
 Here we used the changes in signal energy as well as interconnectivity at different frequency levels to discriminate between seizure  and non-seizure from a data set sourced from PhysioBank \cite{P38Gold:2000}. This data consisted of EEG recordings from subjects with intractable seizures.  
 The combination of both methodologies resulting from the time-frequency representation enabled by using wavelets, has provided a superior performance on a data set consisting of 23 electrodes.  
 Computational results indicated which standard 
statistical features were utilised on the reconstructed signal as well as the Graph theoretic indicators. These features within a highly imbalanced data set are then classified by  a  cost-sensitive decision forest algorithm,   achieved a Recall of 72\% and a f-score of 51\%.  Overall for  our combination method,  while only requiring 131 attributes  and using one second duration epochs, provides a better Recall and f-score than the comparative methods when using the same duration epochs, as shown in Table \ref{table:fscore},

\vspace{-1.25cm}
\begin{IEEEbiographynophoto}
{Paul G. Grant} (M'2022) was born in Hamilton, Vic. Australia. 
This author became  a Member (M) of IEEE in 2022. He received his B.App.Sc degree
\{mathematics\} from Royal Melbourne Institute of Technology, Melbourne, Vic. in 1993; GradDipEc in 1999, MScStud degree\{pure mathematics\} in 2004, MSc degree \{statistics\} in 2010 and GradDipSc  in 2013 from University of New England, Armidale, NSW also a GradDipBiostat in 2009 from University of Sydney, Sydney NSW.
 He is currently 
pursing a PhD degree in Computer Science at Charles Sturt University, Bathurst NSW.
He has been employed in various industrial sectors; Telecommunications, Transport, Insurance and Public Health.
His research interests includes  statistical methods, wavelets, temporal series analysis and data mining.
\end{IEEEbiographynophoto}
\vspace{-1cm}

\begin{IEEEbiographynophoto}{Md Zahidul Islam} is a Professor in Computer Science, in the School of Computing,
Mathematics and Engineering, Charles Sturt
University, Australia. He is serving as the
Director of the Data Science Research Unit
(DSRU), of the Faculty of Business Justice and Behavioural Sciences, Charles Sturt
University, Australia. He received his PhD
degree from the University of Newcastle,
Australia. His main research interests are
in Data Mining, Classification and Clustering
algorithms, Missing value analysis, Outliers detection, Data Cleaning
and Preprocessing, Privacy Preserving Data Mining, Privacy Issues
due to Data Mining on Social Network Users, and Applications of
Data Mining in Real Life.
 
\end{IEEEbiographynophoto}

\end{document}